\documentclass[12pt]{article}%
\usepackage{amsmath}
\usepackage{amsfonts}
\usepackage{amssymb}
\usepackage{graphicx}
\usepackage{amsmath,amsthm,amscd,amsthm,upref,indentfirst}%

\theoremstyle{plain}
\newtheorem{theorem}{Theorem}

\theoremstyle{definition}

\theoremstyle{remark}

\newtheorem{example}[theorem]{Example}
\numberwithin{equation}{section}
\numberwithin{theorem}{section}
\numberwithin{equation}{section}
\normalfont\upshape

\author{\textit{Steven Duplij $ {\!}^{\sf a}$, Gerald A. Goldin $ {\!}^{\sf b}$,
Vladimir M. Shtelen $ {\!}^{\sf c}$}\\[10pt]
$^{\sf a}${\small University of M\"unster, Germany}\\
{\small \texttt{douplii@uni-muenster.de}}\\
$^{\sf b}${\small Departments of Mathematics and Physics}\\
{\small Rutgers University, New Brunswick, NJ USA}\\
{\small \texttt{geraldgoldin@dimacs.rutgers.edu}}\\
$^{\sf c}${\small Department of Mathematics}\\
{\small Rutgers University, New Brunswick, NJ USA}\\
{\small \texttt{shtelen@math.rutgers.edu}}
}

\title{\textbf{On Lagrangian and Non-Lagrangian Conformal-Invariant Nonlinear
Electrodynamics}}
\begin{document}

\maketitle
\thispagestyle{empty}
\begin{abstract}
A general approach is presented to describing nonlinear classical Maxwell electrodynamics
with conformal symmetry. We introduce generalized nonlinear constitutive
equations, expressed in terms of constitutive tensors dependent on
conformal-invariant functionals of the field strengths. This allows a
characterization of Lagrangian and non-Lagrangian theories. We obtain a
general formula for possible Lagrangian densities in nonlinear conformal-invariant
electrodynamics. This generalizes the standard Lagrangian of classical linear
electrodynamics so as to preserve the conformal symmetry.

\end{abstract}

\newpage
\tableofcontents

\vskip 3cm

\section{Introduction}

The conformal invariance of Maxwell's equations in Minkowski spacetime $M^{(4)}$
is well known. Kastrup \cite{kas2} provides a historical review of conformal
symmetry in geometry and physics. Our purpose in the present article is to
develop \textit{nonlinear} electrodynamics with conformal symmetry, both
Lagrangian and non-Lagrangian.

In earlier work \cite{gol/sht1,gol/sht2}, two of us considered general
nonlinear Maxwell and Yang-Mills equations in $M^{(4)}$, satisfying Lorentz
symmetry. The relation between the field strength tensor $F$ and the
displacement tensor $G$ was provided by nonlinear constitutive equations,
which depend only on Lorentz-invariant functionals of the field strengths.
Broadly speaking, such nonlinear (classical) Maxwell equations may describe
electromagnetic fields in matter, or may provide a phenomenological
description of possible fundamental properties of spacetime (e.g., as
described by Born-Infeld or Euler-Kockel Lagrangians). We obtained an explicit
condition distinguishing Lagrangian from non-Lagrangian theories.

This approach was subsequently generalized to include supersymmetric classical
electrodynamics \cite{dup/gol/sht1} and found an application to the null-string
limit of Born-Infeld theory \cite{gol/mav/sza}.

Here we focus on the behavior of the Maxwell fields and spacetime coordinates under
conformal inversion. First we review briefly nonlinear Maxwell theories in
$M^{(4)}$ with Lorentz symmetry. We write down the symmetry of Maxwell's
equations under conformal inversion, and discuss conformal invariant (or
pseudoinvariant) functionals of the field strengths in $M^{(4)}$.

Then we again introduce generalized nonlinear constitutive
equations, expressed in terms of constitutive tensors dependent on
conformal-invariant functionals. This allows
characterization of Lagrangian and non-Lagrangian theories, and leads to a
general formula for possible Lagrangian densities in nonlinear conformal-invariant
electrodynamics. Our results generalize the standard Lagrangian of classical linear
electrodynamics subject to the preservation of conformal symmetry.
The introduced approach differs from other work on conformal invariant nonlinear electrodynamics, e.g. \cite{rig/ven,den/dol/sok}  and references therein.
\section{Conformal Symmetry of Maxwell's Equations}

Recall that the conformal group for the Minkowski $\left(  3+1\right)
$-dimensional spacetime $M^{\left(  4\right)  }$ (described by coordinates
$x^{\mu}$, $\mu=0,1,2,3$) includes spacetime translations, Lorentz boosts, and
dilations, together with the special conformal transformations
\cite{lan/lif2,jackson}. The latter can be obtained as a sequence of
transformations: conformal inversion, translation, and conformal inversion
again. The conformal inversion is given by (summing over repeated indices)%
\begin{equation}
x^{\prime\,\mu}=\frac{x^{\,\mu}}{x_{\nu}x^{\nu}}\,. \label{inversion}%
\end{equation}

A discussion of conformal inversion is provided in \cite{gol/sht/dup2018}.
The special
conformal transformations are%
\begin{equation}
x^{\prime\,\mu}=\frac{(x^{\,\mu}-b^{\,\mu}x_{\nu}x^{\,\nu})}{(1-2b_{\nu
}x^{\,\nu}+b_{\nu}b^{\,\nu}x_{\mu}x^{\,\mu})}\,\,. \label{specialconformal}%
\end{equation}

Next we write the well-known conformal symmetry of electromagnetism \cite{jackson}, with
respect to the transformation (\ref{inversion}).

In covariant notation we have the standard electromagnetic tensor fields%
\begin{equation}
F_{\mu\nu}=\left(  \!\!%
\begin{array}
[c]{cccc}%
0 & \frac{1}{c}\,E_{x} & \frac{1}{c}\,E_{y} & \frac{1}{c}\,E_{z}\\
-\frac{1}{c}\,E_{x} & 0 & -B_{z} & B_{y}\\
-\frac{1}{c}\,E_{y} & B_{z} & 0 & -B_{x}\\
-\frac{1}{c}\,E_{z} & -B_{y} & B_{x} & 0
\end{array}
\!\!\!\right)  ,\,\,G_{\mu\nu}=\left(  \!\!%
\begin{array}
[c]{cccc}%
0 & cD_{x} & cD_{y} & cD_{z}\\
-cD_{x} & 0 & -H_{z} & H_{y}\\
-cD_{y} & H_{z} & 0 & -H_{x}\\
-cD_{z} & -H_{y} & H_{x} & 0
\end{array}
\!\!\!\right)  \!;
\end{equation}
the Hodge dual tensors are $\tilde{F}^{\mu\nu}=\frac{1}{2}\varepsilon^{\mu
\nu\rho\sigma}F_{\rho\sigma}$ and $\tilde{G}^{\mu\nu}=\frac{1}{2}%
\varepsilon^{\mu\nu\rho\sigma}G_{\rho\sigma}$, where $\varepsilon^{\mu\nu
\rho\sigma}$ is the totally antisymmetric Levi-Civita tensor with
$\varepsilon^{0123}=1$. Then Maxwell's equations are
\begin{equation}
\partial_{\mu}\tilde{F}^{\mu\nu}=0,\ \ \ \ \ \ \partial_{\mu}G^{\mu\nu}%
=j^{\nu}, \label{max4}%
\end{equation}
where $j^{\mu}=\left(  c\rho,\mathbf{j}\right)  $ is the $4$-current. The first
equations in (\ref{max4}) imply that one can set $F_{\mu\nu}=\partial_{\mu}A_{\nu}%
-\partial_{\nu}A_{\mu}$, where $A_{\mu}$ is an abelian gauge field; but in general there
is no such representation for $G_{\mu\nu}$. The field strength tensors
$F_{\mu\nu}$ and $\tilde{F}^{\mu\nu}$ are physically observable, in that their
components can be inferred from measurement of the force $\,\mathbf{F}%
=q\left(  \mathbf{E}+\mathbf{v\times B}\right)  $ on a test charge $q$ moving
with velocity $\mathbf{v}$. The relation of the tensors $G_{\mu\nu}$ and
$\tilde{G}^{\mu\nu}$ to observable fields is determined by the properties of
the medium.

Under the inversion (\ref{inversion}) we obtain the coordinate transformations
\cite{dup/gol/sht3}%
\begin{equation}
\partial_{\mu}^{\,\prime}:=\frac{\partial}{{\partial x^{\prime}}^{\mu}}%
=x^{2}\partial_{\mu}-2x_{\mu}(x\cdot\partial)\,, \label{inversionpartial}%
\end{equation}%
\begin{equation}
\Box^{\,\prime}:= \partial_{\mu}^{\,\prime}\partial^{\,\prime\mu} = (x^{2})^{2}\Box-4x^{2}(x\cdot\partial)\,. \label{inversionBox}%
\end{equation}
The transformations of the fields that respect the symmetry are then%
\begin{equation}
A_{\mu}^{\prime}(x^{\prime})=x^{2}A_{\mu}(x)-2x_{\mu}(x^{\alpha}A_{\alpha
}(x))\,, \label{inversionA}%
\end{equation}%
\begin{equation}
F_{\mu\nu}^{\,\prime}(x^{\prime})=(x^{2})^{2}F_{\mu\nu}(x)-2x^{2}x^{\alpha
}(x_{\mu}F_{\alpha\nu}(x)+x_{\nu}F_{\mu\alpha}(x)), \label{inversionF}%
\end{equation}
where $x^{2}=x_{\mu}x^{\mu}$ and $(x\cdot\partial)=x^{\alpha}\partial_{\alpha
}$. Combining this symmetry with that of the Poincar\'{e} transformations and
dilations, we have the well-known symmetry with respect to the conformal
group.

In $M^{(4)}$ we have two fundamental Poincar\'{e}-invariant functionals
\begin{equation}
I_{1}\,=\,\frac{1}{2}F_{\mu\nu}(x)F^{\mu\nu}(x)\,,\quad I_{2}\,=\,-\frac{c}%
{4}{F_{\mu\nu}}(x){{\tilde{F}}}^{\mu\nu}(x)\,. \label{invariantsM}%
\end{equation}
It is known \cite{lan/lif2},
that any other  
Poincar\'{e}-invariant functional can be expressed in terms of $I_{1}$ and $I_{2}$  (see also \cite{esc/urr}). Thus, basing our description of relativistic nonlinear electrodynamics on
$I_1$ and $I_2$ should achieve the maximum level of generality, subject to
writing constitutive equations in the form described in the next section.
Sometimes the functional $I_{2}$ is called a pseudoinvariant, because it
changes sign under spatial reflection (parity) \cite{jackson}. 

Under conformal inversion, $I_{1}$ and $I_{2}$ are no longer individually
invariant. Rather, they transform by%
\begin{equation}
{I_{1}^{\prime}}(x^{\prime})\,=\,\frac{1}{2}{F_{\mu\nu}^{\,\prime}}(x^{\prime
}){({F^{\,\prime})^{\mu\nu}}}(x^{\prime})\,=\,(x^{2})^{4}I_{1}(x)\,,
\label{inversionI1}%
\end{equation}%
\begin{equation}
{I_{2}^{\prime}}(x^{\prime})\,=\,-\frac{c}{4}{F_{\mu\nu}^{\,\prime}}%
(x^{\prime}){({\tilde{F}}^{\,\prime})}^{\mu\nu}(x^{\prime})\,=\,-(x^{2}%
)^{4}I_{2}(x)\,. \label{inversionI2}%
\end{equation}
So the ratio $I_{2}(x)/I_{1}(x)$ is a pseudoinvariant under conformal
inversion%
\begin{equation}
\frac{{I_{2}^{\prime}}(x^{\prime}%
)}{{I_{1}^{\prime}}(x^{\prime})}=-\frac{I_{2}(x)}{I_{1}(x)}  . \label{i21}%
\end{equation}

Because the special conformal transformations involve two inversions, the
functional $I_{2}(x)/I_{1}(x)$ is a true invariant under the special conformal group.

\section[Maxwell's equations and constitutive equations]{Maxwell's equations and nonlinear constitutive equations}

Let us recall that in SI units Maxwell's equations are
\begin{align}
\operatorname{curl}\mathbf{E}  &  =-\dfrac{\partial\mathbf{B}}{\partial
t},\ \ \ \ \ \ \operatorname{div}\mathbf{B}=0,\nonumber\\
\operatorname{curl}\mathbf{H}  &  =\dfrac{\partial\mathbf{D}}{\partial
t}+\mathbf{j},\ \ \ \operatorname{div}\mathbf{D}=\rho, \label{max3a}%
\end{align}
where $\mathbf{j}$ and $\rho$ are current and charge densities \cite{lan/lif2,jackson}. We consider
only flat spacetime, so that the metric is given by the Minkowski metric
tensor is $\eta_{\mu\nu}=\mathrm{diag}\left(  1,-1,-1,-1\right)  $, $x^{\mu
}=\left(  ct,x^{i}\right)  $, $\mu,\nu,\dots\,=\,0,1,2,3$; $i=\,1,2,3$; with
$\left(  \partial_{\mu}\right)  =\left(  \partial\diagup\partial x^{\mu
}\right)  \,=\,\left(  c^{-1}\partial\diagup\partial t,\nabla\right)  $. The
Lorentz invariants (\ref{invariantsM}) in terms of $\mathbf{E},\,\mathbf{B}%
,\,\mathbf{D},\,\mathbf{H}$ are
\begin{equation}
I_{1}=\mathbf{B}^{2}-\dfrac{1}{c^{2}}\mathbf{E}^{2},\ I_{2}=\mathbf{B}%
\cdot\mathbf{E}. \label{cc}%
\end{equation}
The constitutive equations relating $\mathbf{E}$, $\mathbf{B}$, $\mathbf{D}$
and $\mathbf{H}$, taken together Maxwell's equations (\ref{max3a}), determine the symmetry.

General nonlinear constitutive equations with Poincar\'{e} symmetry take the form \cite{fus/sht/ser,fus/tsi,gol/sht1}
\begin{align}
\mathbf{D}  &  =M\left(  I_{1},I_{2}\right)  \mathbf{B}+\dfrac{1}{c^{2}%
}N\left(  I_{1},I_{2}\right)  \mathbf{E},\nonumber\\
\mathbf{H}  &  =N\left(  I_{1},I_{2}\right)  \mathbf{B}-M\left(  I_{1}%
,I_{2}\right)  \mathbf{E}, \label{con1}%
\end{align}
where $M\left(  I_{1},I_{2}\right)  $ and $N\left(  I_{1},I_{2}\right)  $ are
smooth scalar functions of the invariants (\ref{cc}). The constitutive equations in vacuum are $\mathbf{D}=\varepsilon_{0}
\mathbf{E}$ and $\mathbf{B}=\mu_{0}\mathbf{H}$, where $\varepsilon_{0}$ and
$\mu_{0}$ are respectively the permittivity and permeability of empty space, with
$\varepsilon_{0}\mu_{0}=c^{-2}$; (so that $M=0$, and $N=1/\mu_{0}=\varepsilon
_{0}c^{2}$).

In covariant form, the constitutive equations (\ref{con1}) become%
\begin{equation}
G_{\mu\nu}=N\left(  I_{1},I_{2}\right)  F_{\mu\nu}+cM\left(  I_{1}%
,I_{2}\right)  \tilde{F}_{\mu\nu}. \label{gn}%
\end{equation}
Now, because of (\ref{inversionI1})--(\ref{i21}), the
constitutive functionals\ $M$ and $N$ in a system with conformal symmetry can depend \textit{only} on the ratio $I_{2}(x)/I_{1}(x)$. Then let us write $M\left(  I_{1},I_{2}\right)  =\mathcal{M}\left(
{u}\right)  $ and $N\left(  I_{1},I_{2}\right)  =\mathcal{N}\left(
{u}\right)  $ \cite{fus/tsi,fus/sht/ser}. For convenience, we denote
\begin{equation}
I_{1}=\dfrac{1}{2}F_{\mu\nu}F^{\mu\nu}\equiv X,\ \ \ \ \ \ \ \ I_{2}%
=-\dfrac{c}{4}F_{\mu\nu}\tilde{F}^{\mu\nu}\equiv Y. \label{xy}%
\end{equation}
Then we rewrite the constitutive equations (\ref{con1}) in the covariant form%
\begin{equation}
G_{\mu\nu}^{\left(  \mathrm{conformal}\right)  }\equiv G_{\mu\nu}%
=\mathcal{N}\left(  u\right)  F_{\mu\nu}+c\mathcal{M}\left(  u\right)
\tilde{F}_{\mu\nu}, \label{con2}%
\end{equation}
where  \begin{equation}u=\frac{Y}{cX}\label{u}
\end{equation} is a dimensionless conformal invariant.

\section{Lagrangian and non-Lagrangian theories}

When the constitutive equations are nonlinear, one may have either a Lagrangian or a
non-Lagrangian theory. In the first case,
the equations of motion are derived from a Lagrangian density $L\left(  X,Y\right)$.
Then the constitutive equations are
\begin{equation}
G_{\mu\nu}=-\frac{\partial L}{\partial F^{\mu\nu}} = -2\dfrac{\partial L\left(  X,Y\right)  }{\partial X} F_{\mu\nu}
+c\dfrac{\partial L\left(  X,Y\right)  }{\partial Y}\tilde{F}_{\mu\nu}.
\label{gf}%
\end{equation}
The functions $M$ and $N$ in (\ref{gn}) may then be written%
\begin{align}
N\left(  X,Y\right)   &  =\frac{1}{\mu}+N_{L}\left(  X,Y\right)  =-2\dfrac{\partial
L\left(  X,Y\right)  }{\partial X},\label{nm}\\
M\left(  X,Y\right)   &  =M_{L}\left(  X,Y\right)  =
\dfrac{\partial L\left(  X,Y\right)  }{\partial Y}. \label{nm1}%
\end{align}

We singled out the constant $1/\mu$ in (\ref{nm}) so that the choice $N_{L}\left(
X,Y\right)  =M_{L}\left(  X,Y\right)  =0$ and $\mu=\mu_0$ yields the standard Lagrangian of
linear electrodynamics in vacuo, $L_{0}\left(  X,Y\right)  =-(1/2\mu_0)X$ \cite{jackson}. Then a
necessary and sufficient condition for the theory to be Lagrangian is that the cross-derivatives be equal; i.e.,%
\begin{equation}
-\frac{1}{2} \dfrac{\partial N\left(  X,Y\right)  }{\partial Y}=\dfrac{\partial
M\left(  X,Y\right)  }{\partial X}\,, \label{nmy}%
\end{equation}

\medskip
\noindent
where both sides must equal $\partial^{2}L\left(  X,Y\right)/{\partial X\partial Y}$.
If $M$ and $N$ in (\ref{gn}) violate (\ref{nmy}), then the theory cannot be
Lagrangian \cite{gol/sht2}.

\section[Conformal-invariant Lagrangian NED]{Conformal-invariant Lagrangian nonlinear electrodynamics}

Now let us consider conformal-invariant nonlinear electrodynamics, with the
constitutive equations (\ref{con2}). If the theory is Lagrangian, we use the subscript $L$ and
write $M_L(X,Y) = \mathcal{M}_{L}(u)$ and $N_L(X,Y) = \mathcal{N}_{L}(u)$ (so that in the Lagrangian case $\mathcal{N} = [1/\mu]+ \mathcal{N}_L$ and $\mathcal{M} = \mathcal{M}_L)$, to remind us of the dependence of these functionals on the specific choice of Lagrangian $L$. Then (\ref{nmy}) takes the form%
\begin{equation}
\frac{d\mathcal{N}_{L}\left(  u\right)  }{du}-2cu\frac{d\mathcal{M}_{L}\left(
u\right)  }{du}=0,\ \ \ \ u=\frac{Y}{cX}. \label{uc}%
\end{equation}
Integrating (\ref{uc}), we can express $\mathcal{N}_{L}\left(  u\right)$ as%
\begin{equation}
\mathcal{N}_{L}\left(  u\right)  =2c\int  u\frac{d\mathcal{M}_{L}\left(
u\right)  }{du}du=2cu\mathcal{M}_{L}\left(  u\right)  -2c\int\mathcal{M}%
_{L}\left(  u\right)  du,
\end{equation}

\smallskip
\noindent
with the constant of integration subsumed into the term $1/\mu$.

Thus we obtain one form of the general constitutive equations respecting conformal symmetry,%
\begin{equation}
G_{\mu\nu}=\left(  \frac{1}{\mu}+2uc\mathcal{M}_{L}\left(  u\right)  -2c\int\mathcal{M}%
_{L}\left(  u\right)  du\right)  F_{\mu\nu}+c\mathcal{M}_{L}\left(  u\right)
\tilde{F}_{\mu\nu}\,. \label{gnf}%
\end{equation}
The constitutive equations as represented by (\ref{gnf}) depend on one arbitrary functional $\mathcal{M}_{L}\left(  u\right)$, a function of the ratio of relativistic invariants $u= Y/cX$.

There may be a constant term in $\mathcal{M}_L$. However, as was earlier remarked in \cite{gol/sht1}, adding a constant $\kappa$ to $M$ (i.e., to $\mathcal{M}_L$) does not change the observable physics. Indeed, referring back to (\ref{max3a})-(\ref{con1}), a result of adding $\kappa$ to $M$ is to add a term $\kappa \mathbf{B}$ to $\mathbf{D}$. But since $\mathrm{div} \, \mathbf{B}$ = 0, the value of $\rho = \mathrm{div} \,\mathbf {D}$ is unchanged. Likewise, a term $ - \kappa \mathbf{E}$ is added to $\mathbf{H}$. But the resulting term in the equation for $\mathbf{j}$ is offset by the term that was added to $\mathbf{D}$. Hence the system $\mathbf{E}$, $\mathbf{B}$, $\rho$, $\mathbf{j}$ is unaffected by $\kappa$; but it is these fields which describe all the observable forces produced by and acting on electric charges and currents.

Equivalently to (\ref{uc}), one may write
\begin{equation}
\mathcal{M}_L = \frac{1}{2c} \int \frac{1}{u} \frac{d \mathcal{N}_L}{du}\,du
\end{equation}
and (\ref{gnf}) becomes
\begin{equation}
G_{\mu\nu}=\left(  \frac{1}{\mu}+\mathcal{N}_{L}\left(  u\right)\right) F_{\mu\nu}
+ \left(\frac{1}{2}\int \frac{1}{u} \frac{d \mathcal{N}_L}{du}\,du \right)
\tilde{F}_{\mu\nu}\,,%
\end{equation}
in which $\mathcal{N}_{L}\left(  u\right)$ is taken to be the arbitrary function
of $u$.

The general Lagrangian density $\,L = L_{\mathrm{nonlin}}^{\mathrm{conform}}\left(X,Y\right)\,$ for conformal-invariant nonlinear electrodynamics can
now be written in several equivalent forms; e.g.,
\begin{equation}
L = L_0 + cX\int \mathcal{M}_L(u) du = L_0 + Y\left(\frac{1}{u}\int \mathcal{M}_L(u) du \right),
\label{LnlM}
\end{equation}
%
%
or
\begin{equation}
L = L_0 + Y\left(\frac{1}{2c}\int \frac{1}{u^2}\,\mathcal{N}_L(u) du\right), \label{LnlN}
\end{equation}
where $L_{0}=-(1/2\mu) X$ describes standard linear electrodynamics with
$\mu=\mu_0$ and $G_{\mu\nu}= ({1}/\mu_0)F_{\mu\nu}$. 
As noted after (\ref{gnf}), if $\mathcal{M}_{L}\left(  u\right) = \kappa$ (a constant), then $$L_{\mathrm{lin}} = -\frac{1}{2\mu} X + \kappa Y$$ describes electrodynamics physically equivalent to that described by $L_0$. Otherwise, the general conformal-invariant electrodynamics described by (\ref{LnlM}) or (\ref{LnlN}) is nonlinear.

The following examples illustrate some of the many possibilities.

\begin{example}
Let $\mathcal{M}_{L}\left(  u\right)  =\lambda_{1} u$, where the coefficient $\lambda_{1}$ (with the dimensionality of $\epsilon_{0}c$) controls the magnitude of the nonlinearity. Then $\mathcal{N}_{L}\left(u\right)  = \lambda_{1} cu^2$, and
\begin{equation}
L   =-\frac{1}{2\mu}X+\lambda_{1}\frac{Y^{2}}{2cX}\,,
\end{equation}

\smallskip

\begin{equation}
G_{\mu\nu}   =\left(  \frac{1}{\mu}+\lambda_{1}\frac{Y^{2}}{cX^{2}}\right)  F_{\mu\nu}+\lambda_{1} \frac{Y}{X}\tilde{F}_{\mu\nu}\,.
\end{equation}

\end{example}

\begin{example}
More generally, let $\mathcal{M}_{L}\left(  u\right)  =\lambda_{n} u^n$, $n \neq -1$ (where again, $\lambda_{n}$ has the dimensionality of $\epsilon_{0}c$). Then $\mathcal{N}_{L}\left(u\right)  = \lambda_{n}\, [2cn/(n+1)]\, u^{n+1}$; and
\begin{equation}
L   =-\frac{1}{2\mu}X+\lambda_{n}\,\,\frac{1}{(n+1)}\,\,\frac{Y^{n+1}}{c^nX^n}\,,
\end{equation}

\smallskip

\begin{equation}
G_{\mu\nu}   =\left(  \frac{1}{\mu}+\lambda_{n}\,\,\frac{2n}{(n+1)}\,\,\frac{Y^{n+1}}{c^nX^{n+1}}\right)  F_{\mu\nu}+\lambda_{n} \frac{Y^n}{c^{n-1}X^n}\tilde{F}_{\mu\nu}\,.
\end{equation}

\noindent
Note that if $n < 0$, the model is singular when $Y = 0$; i.e., when $\mathbf{B}\cdot\mathbf{E} = 0$.

\end{example}

\begin{example}
Let $\mathcal{N}_{L}\left(  u\right)  =\alpha u$ (where $\alpha$ has the dimensionality of $1/\mu_0$ or $\epsilon_0c^2$). Then $\mathcal{M}_L(u) = (\alpha/2c)\ln |u|$, and we obtain
\begin{equation}
L   =-\frac{1}{2\mu}X+ \alpha \frac{Y}{2c} \ln\left|\frac{Y}{cX}\right|\,,
\end{equation}

\smallskip

\begin{equation}
G_{\mu\nu} =
\left(\frac{1}{\mu}+\alpha\frac{Y}{cX}\right)F_{\mu\nu}+\frac{\alpha}{2}\ln\left|\frac{Y}{cX}\right|%
\tilde{F}_{\mu\nu}\,.
\end{equation}

\noindent
This model also is singular when $u = 0$; i.e., when $\mathbf{B}\cdot\mathbf{E} = 0$.
\end{example}

\begin{example}
We may let $\mathcal{M}_{L}\left(  u\right)  =\lambda \sin bu$, where $b\neq0$ is an additional dimensionless parameter. Then $\mathcal{N}_{L}\left(u\right)  = 2\lambda c\, (u \sin bu + b^{-1} \cos bu)$; and now,

\begin{equation}
L   =-\frac{1}{2\mu}X - \lambda\,\,\frac{c}{b}\,X \cos \left(\frac{b\,Y}{cX}\right)\,.
\end{equation}

\smallskip

\begin{equation}
G_{\mu\nu}   =\left[  \frac{1}{\mu} + 2\lambda \, \left(\frac{Y}{X}\right) \sin \left(\frac{b\,Y}{cX}\right) + \frac{c}{b}\cos \left(\frac{b\,Y}{cX}\right) \right] F_{\mu\nu}
 + \lambda c \sin \left(\frac{b\,Y}{cX}\right)\,\tilde{F}_{\mu\nu}\,.
\end{equation}

\end{example}

One can obtain similar equations for the examples $\mathcal{M}_{L}\left(  u\right)  =\lambda \cos bu$, $\mathcal{M}_{L}\left(  u\right)  =\lambda \sinh bu$, and $\mathcal{M}_{L}\left(  u\right)  =\lambda \cosh bu$.

\section{Conclusion}

We have described an approach to nonlinear classical Maxwell
electrodynamics with conformal symmetry, based on generalized
constitutive equations. The latter are expressed in terms of constitutive tensors
that depend on conformal-invariant functionals of the field strengths. Our
general description includes both Lagrangian and non-Lagrangian theories. The latter would of course include nonconservative, or dissipative systems.

A straightforward criterion distinguishes the Lagrangian case, which leads to
a general formula for the Lagrangian density. We present several examples, illustrating a variety of possibilities for nonlinear conformal-invariant electrodynamics. These may occur as nonlinear perturbations of the usual, linear Maxwell theory.

By requiring conformal symmetry, we thus distinguish a particular class of nonlinear Lagrangian theories. Some choices may be candidates for the phenomenological description of conservative electrodynamics in the presence of matter, or even for describing possible fundamental properties of spacetime. We remark here that the Born-Infeld and Euler-Kockel Lagrangians are not of the form of (\ref{LnlM}) or (\ref{LnlN}), and thus break the conformal symmetry.

Elsewhere we have suggested further generalization based on the well-known identification of conformally compactified Minkowski spacetime with the projective light cone in a $(4+2)$-dimensional spacetime $Y^{(6)}$ \cite{dup/gol/sht3}. The conformal symmetry is then expressed through (ordinary and hyperbolic) rotations in $Y^{(6)}$. Writing the $6$-dimensional analog of Maxwell's equations, there are now two independent conformal-invariant functionals of the field strengths. Dimensional reduction then permits one to recover additional possibilities in Minkowski spacetime $M^{(4)}$. This is a topic of our ongoing research.


\end{document}